\documentclass[runningheads]{cl2emult}

\usepackage{makeidx}  % allows index generation
\usepackage{graphicx} % standard LaTeX graphics tool
\usepackage{subeqnar} % subnumbers individual equations
\usepackage{multicol} % used for the two-column index
\usepackage{cropmark} % cropmarks for pages without
\usepackage{eso}      % placeholder for figures
\makeindex            % used for the subject index
\begin{document}
\title*{Black Hole Formation and Gamma Ray Bursts}
\toctitle{Black Hole Formation and Gamma Ray Bursts}
% allows explicit linebreak for the table of content
%
%
\titlerunning{Black Hole Formation and Gamma Ray Bursts}
% allows abbreviation of title, if the full title is too long
% to fit in the running head
%
\author{Remo Ruffini}
\authorrunning{Remo Ruffini}
% if there are more than two authors,
% please abbreviate author list for running head
%
%
\institute{I.C.R.A.--International Center for Relativistic Astrophysics
and
Physics Department, University of Rome ``La Sapienza",\\
 I-00185 Rome, Italy}

\maketitle              % typesets the title of the contribution

\begin{abstract}
Recent work on the dyadosphere of a black hole is reviewed with special emphasis on the explanation of gamma ray bursts. A change of paradigm in the observations of black holes is presented. 
\end{abstract}

\section{Introduction}

\begin{figure}
\centering
\includegraphics[width=0.6\textwidth]{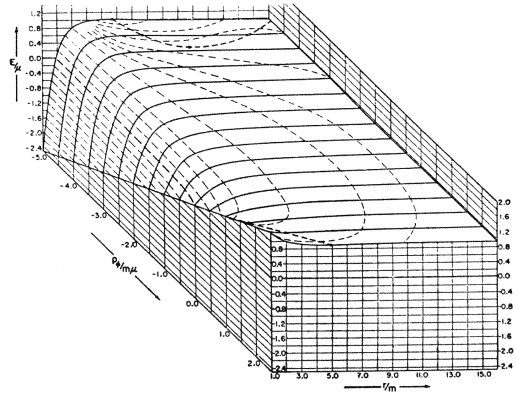}
\caption[]{``Effective potential" around a Kerr black hole, see Ruffini and Wheeler 1971}
\label{poten}
\end{figure}

An ``effective
potential" technique had been used very successfully by Carl St\o rmer in the
1930s in studying the trajectories of cosmic rays in the Earth's magnetic field
(St\o rmer 1934).
In the fall of 1967 Brandon Carter visited Princeton and presented his
remarkable mathematical work leading to the separability of the
Hamilton-Jacobi equations for the trajectories of charged particles in the
field of a Kerr-Newmann geometry (Carter 1968). This visit had a profound impact on our small
group working with John Wheeler on the physics of gravitational collapse.
Indeed it was Johnny who had the idea to use the St\o rmer ``effective
potential" technique in order to obtain physical consequences from the
set of first order differential equations obtained by Carter. I still remember the $2m\times 2m$ grid plot of the effective potential for particles
around a Kerr metric I prepared which finally appeared in print (Ruffini
and Wheeler 1971) and (Rees, Ruffini and Wheeler 1973,1974); see Fig.(\ref{poten}). 
%---------------------------------------------figure potential--------------------%
%---------------------------------------------figure potential--------------------% 
From this work came the celebrated result of the maximum binding energy of  $1 - {1 \over \sqrt{3}}\sim42\%$ 
for corotating orbits and 
$1-{5\over 3\sqrt{3}}\sim 3.78\%$ for counter-rotating orbits
 in the Kerr geometry. We were very pleased to be associated with Brandon Carter in a ``gold
medal" award for this work presented by Yevgeny Lifshitz: in the last edition of volume 2 of the Landau
and Lifshitz series ({\it The Classical Theory of Fields\/}), both Brandon's work and my own work with Wheeler were proposed as named exercises for bright students!
During this meeting it was also gratifying to hear in the talks of Rashid Sunyaev and others that these
results have become the object of direct observations in X-ray sources.

The ``uniqueness theorem" stating that black holes can only be characterized
by their mass-energy $E$, charge $Q$ and angular momentum $L$ had been advanced
in our article ``Introducing the Black Hole"(Ruffini and Wheeler 1971) with its
very unconventional figure in which TV sets, bread, flowers and other objects lose their
characteristic features and merge in the process of gravitational collapse into the three
fundamental parameters of a black hole. That picture became the object of a great deal of 
lighthearted discussion in the physics community. A proof of this uniqueness theorem, satisfactory for most cases of astrophysical interest, has been obtained after twenty five years of meticulous mathematical work (see e.g., Regge and Wheeler (1957), Zerilli (1970,1974), Teukolsky (1973), C.H. Lee  (1976,1981), Chandrasekhar (1976,1983)). However the proof  still presents some outstanding difficulties in
its most general form. Possibly some progress will be reached in the near future
with the help of computer algebraic manipulation techniques to overcome the
extremely difficult mathematical calculations (see e.g., Cruciani 1999).

The ``maximum mass of a neutron star" was the subject of the thesis
of Clifford Rhoades, my second graduate student at Princeton. A criteria was found there to overcome fundamental unknowns about the behaviour of matter at supranuclear densities by establishing an absolute upper limit to the neutron star mass based only on general relativity, causality and the behaviour of matter at nuclear and subnuclear densities.   This work, presented at the 1972 Les Houches summer School (B. and C. de Witt 1973), only appeared after a prolongued debate (see the reception and publication dates!)(Rhoades and Ruffini 1974).

\begin{itemize}

\item 
The ``black hole uniqueness theorem", implying the axial symmetry of the
configuration and the absence of regular pulsations from black holes,
\item 
 the ``effective potential technique", determining the efficiency of the energy emission in the accretion process, and 
\item 
the ``upper limit on the maximum mass of a neutron star" discriminating between 
an unmagnetized neutron star and a black hole
\end{itemize}
were the three essential components in establishing the paradigm
for the identification of the first black hole in Cygnus X1 (Leach and Ruffini 1973).
These results were also presented in a widely attended session chaired 
by John Wheeler at the 1972 Texas Symposium in New York, extensively reported by the New York Times. The New York Academy of Sciences which hosted the symposium had just awarded me their prestigious Cressy Morrison Award for my work on neutron stars and black holes. Much to their dismay I never wrote the paper for the proceedings since it coincided with the one submitted for publication (Leach and Ruffini 1973). 

The definition of the paradigm did not
come easily but slowly matured after innumerable discussions, mainly on
the phone, both with Riccardo Giacconi and Herb Gursky. I still remember an irate
professor of the Physics Department at Princeton pointing publicly to my outrageous phone
bill of \$274 for one month, at the time considered scandalous, due to my
frequent calls to the Smithsonian, and a much more
relaxed and sympathetic attitude about this situation by the department chairman, Murph Goldberger. 
This work was finally summarized in two books: one with Herbert Gursky (Gursky and Ruffini 1975), following the 1973 AAAS Annual Meeting in San Francisco, and the second with Riccardo Giacconi (Giacconi and Ruffini 1975) following the 1975 LXV Enrico Fermi Summer School (see also the proceedings of the 1973 Solvay Conference). 

\begin{figure}
\centering
\includegraphics[width=0.6\textwidth]{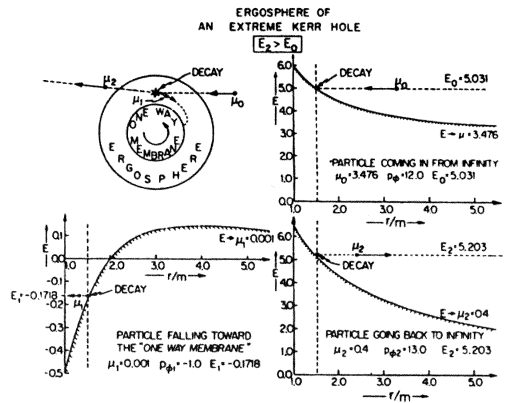}
\caption[]{(Reproduced from Ruffini and Wheeler with their kind permission.) Decay of a particle of rest-plus-kinetic energy $E_\circ$
     into a particle which is captured by the black hole with positive
     energy as judged locally, but negative energy $E_1$ as judged from infinity, together with a particle of rest-plus-kinetic energy $E_2>E_\circ$ which escapes to infinity. The cross-hatched curves give the effective potential (gravitational plus centrifugal) defined by the solution $E$ of Eq.(2) for constant values of $p_\phi$ and $\mu$ (figure and caption
reproduced from Christodoulou 1970).}
\label{pic1}
\end{figure}

The effective potential technique, see Figure (\ref{pic1}), was also essential in order to
explore a suggestion, presented by Penrose at the first meeting of the European
Physical Society in Florence in 1969, that rotational energy could be extracted from black holes.
%---------------------------------------------figure 1--------------------%
%---------------------------------------------figure 2--------------------%
The first specific example of such an energy extraction process by a gedanken experiment was given in Ruffini and Wheeler (1970), see Figure (\ref{pic1}),
and later by Floyd and Penrose (1971).
The reason for showing this figure here is not just to recall the first explicit computation and the introduction of the ``ergosphere", but to emphasize how contrived and difficult such a mechanism can be: it can only work for very special parameters and is in general associated with a reduction of the rest mass of the particle involved in the process. To slow down the rotation of a black hole and to increase its horizon by  accretion of counter-rotating particles is almost trivial, but to extract the rotational energy from a black hole by a reversible transformation in the sense of  Christodoulou and Ruffini (1971), namely to slow down the black hole {\it and} keep its surface area constant, is extremely difficult, as also clearly pointed out by the example in Figure (\ref{pic1}). 

The above gedanken experiments, extended as well to electromagnetic interactions, became very relevant not for their direct astrophysical significance but because they gave the tool for testing the physics and identifying the general mass-energy formula for black holes (Christodoulou and Ruffini 1971): 
\begin{eqnarray}
E^2&=&M^2c^4=\left(M_{\rm ir}c^2 + {Q^2\over2\rho_+}\right)^2+{L^2c^2\over \rho_+^2},\label{em}\\
S&=& 4\pi \rho_+^2=4\pi(r_+^2+{L^2\over c^2M^2})=16\pi\left({G^2\over c^4}\right) M^2_{\rm ir},
\label{sa}
\end{eqnarray}
with
\begin{equation}
{1\over \rho_+^4}\left({G^2\over c^8}\right)\left( Q^4+4L^2c^2\right)\leq 1,
\label{s1}
\end{equation}
where $M_{\rm ir}$ is the irreducible mass, $r_{+}$ is the horizon radius, $\rho_+$ is the quasi-spheroidal cylindrical coordinate of the horizon evaluated at the equatorial plane,
$S$ is the horizon surface area, and extreme black holes satisfy the equality in eq.~(\ref{s1}). The crucial point is that transformations at constant surface area of the black hole, namely reversible transformations, can release an energy up to 29\% of the mass-energy of an extremal rotating black hole and up to 50\% of the mass-energy of an extremely magnetized and charged black hole. Since my Les Houches lectures ``On the energetics of black holes" (B.C. De Witt 1973), one of my main research goals has been to identify an astrophysical setting where the extractable mass-energy of the black hole could manifest itself: I give reasons below why I think that gamma ray bursts (GRBs) are outstanding candidates for observing this extraction process.

\section{the dyadosphere}

At the time of the AAAS meeting in San Francisco (Gursky and Ruffini 1975), we had heard about the  observations of the military Vela satellites which had just been unclassified and we asked Ian B. Strong to report for the first time in a public meeting on gamma ray bursts (Strong 1975). Since those days thousands of publications have appeared on the subject, most irrelevant. One of the
reasons for this is that the basic energetic requirements for GRBs have become clear only recently. The observations of the Compton satellite, through thousands of GRB observations, clearly pointed to the isotropic distribution of these sources in the sky. However, it was only with the very unexpected and fortuitous observations of the Beppo-SAX satellite that the existence of a long lasting afterglow of these sources was identified: this has led to the determination of a much more accurate position for these sources in the sky, which permitted for the first time their optical and radio identification, which in turn has led to the determination of their cosmological distances and to their paramount energetic requirements (see e.g., Frontera and Piro 1999  and references therein). The very fortunate interaction and resonance between X-ray, optical and radio astronomy which in the seventies allowed the maturing of the physics and astrophysics of neutron stars and black holes (see e.g. Giacconi and Ruffini 1977) promises to be active again today in unravelling the physics and astrophysics of the gamma ray burst sources.
	
In 1975, following the work on the energetics of black holes (Christodoulou and Ruffini 1971),
we pointed out (Damour and Ruffini, 1975) the existence of the vacuum polarization process {\it a' la} Heisenberg-Euler-Schwinger (Heisenberg and Euler 1931, Schwinger 1951) around black holes endowed with electromagnetic structure (EMBHs). Such a process can only occur for EMBHs of mass smaller then $7.2\cdot 10^{6}M_\odot$. The basic energetics implications were contained in Table~1 of that paper (Damour and Ruffini, 1975), where it was also shown that this process is almost reversible in the sense introduced by Christodoulou and Ruffini (1971) and that it extract the mass energy of an EMBH very efficiently. We also pointed out that this vacuum polarization process around an EMBH offered a natural mechanism for explaining GRBs.
 
The recent optical observations of GRBs (see e.g. Kulkarni {\it et. al.} 1998), pointing clearly to their cosmological origin and their enormous energy requirements, have convinced us to return to our earlier work in defining more accurately the region of pair creation around an EMBH. This has led to the new concept of the dyadosphere of an EMBH (named for the Greek word for pair) and to the concept of a plasma-electromagnetic (PEM) pulse and its evolution which can generate signals with the characteristic features of a GRB.
I am proposing and giving reasons to support the claim that with gamma ray bursts, we are witnessing for the first time the moment of gravitational collapse to a black hole in real time. Even more importantly, the tremendous energies involved in the energetics of these sources clearly point to the necessity for and give the  opportunity to use the extractable energy of black holes as an energy source for these objects as in Eqs.~(1)--(3) above.

Various models have been proposed in order to tap the rotational energy of black holes by processes of relativistic magnetohydrodynamics (see e.g., Ruffini and Wilson (1975) and \cite{dhrw}). It should be expected, however, that these processes are relevant over the long time scales characteristic of accretion processes. In the present case of gamma ray bursts a sudden mechanism appears to be at work on time scales shorter than a second for depositing the entire energy in the fireball at the moment of the triggering process of the burst, similar to the vacuum polarization process introduced in (Damour  and Ruffini, 1975).
The fundamental new points we have found re-examining this work can be simply summarized, see (Preparata, Ruffini and Xue 1998a,b) for details:

\begin{itemize}

\item 
The vacuum polarization process can  occur in an extended region around the black hole called the dyadosphere, extending from the horizon radius $r_+$ out to the dyadosphere radius $r_{ds}$. Only black holes with a mass larger than the upper limit of a neutron star and up to a maximum mass of $7.2\cdot 10^{6}M_\odot$ can have a dyadosphere.

\item 
The efficiency of transforming the mass-energy of a black hole into particle-antiparticle pairs outside the horizon can approach 100\%, for black holes in the above mass range. 

\item 
The pair created are mainly positron-electron pairs and their number is much larger than the quantity $Q/e$ one would have naively expected on the grounds of qualitative considerations. It is actually given by $N_{\rm pairs}\sim{Q\over e}{r_{ds}\over \hbar/mc}$, where $m$ and $e$ are respectively  the electron mass and charge.  The energy of the pairs and consequently the  emission of the associated electromagnetic radiation peaks in the gamma X-ray region, as a function of the black hole mass.

\end{itemize}

Let us now recall the main results on the dyadosphere obtained in (Preparata, Ruffini and Xue 1998a,b). Although the general considerations presented by Damour and Ruffini (1975) did refer to a Kerr-Newmann field with axial symmetry about the rotation axis, for simplicity, we have considered the case of a nonrotating Reissner-Nordstrom EMBH to illustrate the basic gravitational-electrodynamical process. The dyadosphere then lies between the radius 
\begin{equation}
r_{\rm ds}=\left({\hbar\over mc}\right)^{1\over2}
\left({GM\over c^2}\right)^{1\over2} 
\left({m_{\rm p}\over m}\right)^{1\over2}
\left({e\over q_{\rm p}}\right)^{1\over2}
\left({Q\over\sqrt{G} M}\right)^{1\over2}.
\label{rc}
\end{equation} 
and the horizon radius 
\begin{equation}
r_{+}={GM\over c^2}\left[1+\sqrt{1-{Q^2\over GM^2}}\right].
\label{r+}
\end{equation}
The number density of pairs created in the dyadosphere is 
\begin{equation}
N_{e^+e^-}\simeq {Q-Q_c\over e}\left[1+{
(r_{ds}-r_+)\over {\hbar\over mc}}\right] \ ,
\label{n}
\end{equation}
where $Q_c=4\pi r_+^2{m^2c^3\over \hbar e}$. The total energy of pairs, converted from the static electric energy,
deposited within the dyadosphere is then
\begin{equation}
E^{\rm tot}_{e^+e^-}={1\over2}{Q^2\over r_+}(1-{r_+\over r_{\rm ds}})(1-
\left({r_+\over r_{\rm ds}}\right)^2) ~.
\label{tee}
\end{equation}

\section{The PEM pulse}

The analysis of the radially resolved evolution of the energy
deposited within the $e^+e^-$-pair and photon plasma fluid created
in the dyadosphere of an EMBH is much more complex then we had initially anticipated. Explaining our first attempt to Jim Wilson led him to prononce the Salomonic sentence `` Remo, your bomb will not kill any one!" Some basic ingredients well known to Livermore scientists were missing. We decided to join forces and propose a new collaboration with the Livermore group renewing the successful collaboration with Jim of 1974 (Ruffini and Wilson 1975). We proceeded in parallel: in Rome with simple almost analytic models to  then be  validated by the Livermore codes (Wilson, Salmonson and Mathews 1997,1998).

In Wilson (1975,1977), a black hole charge of the order $10 \%$ was
formed. Thus we assumed a Reissner-Nordstrom black hole with charge $Q = 0.1 Q_{max}$, where $Q_{max}=\sqrt{G}M$. For the evolution we assumed the relativistic hydrodynamic equations, for details see ref.~\cite{rswx}.
We assumed the plasma fluid of $e^+e^-$-pairs, photons and baryons to be a simple perfect fluid in the curved spacetime. The baryon-number and energy-momentum conservation laws are 
\begin{eqnarray}
(n_B U^\mu)_{;\mu}&=&(n_BU^t)_{,t}+{1\over r^2}(r^2 n_BU^r)_{,r}= 0~,
\label{contin}\\
(T_\mu^\sigma)_{;\sigma}&=&0,
\label{contine}
\end{eqnarray}
and the rate equation: 
\begin{equation}
(n_{e^\pm}U^\mu)_{;\mu}=\overline{\sigma v} \left[n_{e^-}(T)n_{e^+}(T) - n_{e^-}n_{e^+}\right] ~,
\label{econtin}
\end{equation}
where $U^\mu$ is the four-velocity of the plasma fluid, $n_B$ the proper baryon-number density, $n_{e^\pm}$ are the proper densities of electrons and positrons($e^\pm$), $\sigma$ is the mean pair annihilation-creation cross-section, $v$ is the thermal velocity of $e^\pm$, and $n_{e^\pm}(T)$ are the proper number-densities of $e^\pm$ at an appropriate equilibrium temperature $T$. The calculations are continued until the plasma fluid
expands, cools and the $e^+e^-$ pairs recombine and the system becomes optically thin.

The results of the Livermore computer code are compared and contrasted with three almost analytical models: 
(i) spherical model: the radial component of
four-velocity is of the form $U(r)=U{r\over {\cal R}}$, where $U$ is the four-velocity at
the surface (${\cal R}$) of the plasma, similar to a portion of a Friedmann model 
(ii) slab 1: $U(r)=U_r={\rm
const.}$, an expanding slab  with constant width ${\cal D}= R_\circ$ in
the coordinate frame in which the plasma is moving; 
(iii) slab 2: 
an expanding slab with constant width $R_2-R_1=R_\circ$ in the comoving frame of the plasma. 

We compute the relativistic Lorentz factor $\gamma$ of the expanding
$e^+e^-$ pair and photon plasma.
%---------------------------------------------figure 2--------------------%
\begin{figure}
\centering
\includegraphics[width=.5\textwidth]{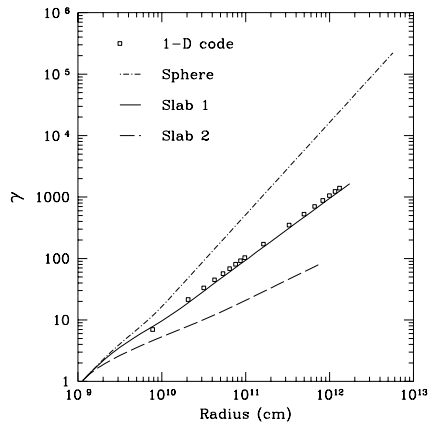}
\caption[]{Lorentz $\gamma$ as a function of radius.
Three models for the expansion pattern of the $e^+e^-$ pair plasma are
compared with the results of the one dimensional hydrodynamic code for
a $1000 M_\odot$ black hole with charge $Q = 0.1 Q_{max}$.  The 1-D
code has an expansion 
pattern that strongly resembles that of a shell
with constant coordinate thickness.}
\label{pic2}
\end{figure}
%---------------------------------------------figure 2--------------------%
In Figure (\ref{pic2}) we see a comparison of the
Lorentz factor of the expanding fluid as a function of radius for all
the models. We can see that the one-dimensional code (only a few 
significant points 
are plotted) matches the
expansion pattern of a shell of constant coordinate thickness.

In analogy with the notorious electromagnetic radiation EM  pulse  of some explosive events, we called this relativistic counterpart 
of an expanding pair electromagnetic radiation shell a PEM pulse.
%---------------------------------------------figure 3--------------------%
\begin{figure}
\centering
\includegraphics[width=.5\textwidth]{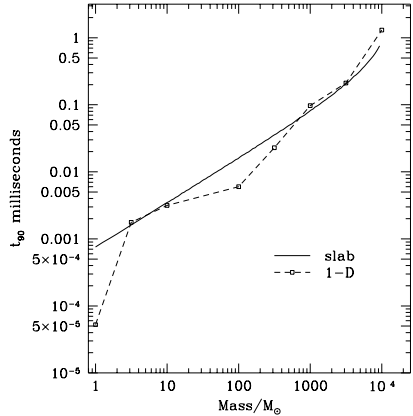}
\caption[]{The duration of the emission at decoupling is
represented by $t_{90}$ plotted over a range of black hole masses.}
\label{pic3}
\end{figure}
%---------------------------------------------figure 3--------------------%
In Figure (\ref{pic3}) we plot correspondingly the time $t_{90}$ over which 90\% of the emission is received from a PEM pulse reaching transparency, as a function of the black hole mass, details given in reference\cite{rswx}. These theoretical predictions 
can be compared and contrasted with the observations. 

\section{Conclusions}

It is well known that pulsars originate their energy from the rotational energy of neutron stars, which gave the evidence for the discovery of neutron stars. Binary X-ray sources originate their energy from the deep relativistic potential well of neutron stars and black holes, and gave the evidence for the discovery of black holes in our galaxy with Cignus X1. We propose that the gamma ray bursts originate their energy from the 
mass-energy of black holes.

The vacuum polarization process we consider can occur in two very distinct regimes: in the collapse of systems leading to black holes of a few solar masses, and in the collapse of very large black holes in the range $10^{3}$ to $10^{6} M_\odot$. While the mechanism of formation for the systems of the first type is well understood, further work is left to be done in understanding the astrophysical settings leading to the collapse of very large EMBHs. Such $10^{3}$ to $10^{6} M_\odot$ black holes should be considered as ``seed black holes" leading by subsequent process of accretion to active galactic nuclei and quasars.

By refining the theoretical models we should be able to retrace the basic parameters of EMBHs from the timing and energy spectrum of GRBs.

Further work is directed toward:
\begin{itemize}
\item 
studying the interaction of the PEM pulse with the baryonic matter of the remnant;
\item 
generalizing our treatment to the rotating case leading to the breakdown of spherical symmetry;
\item 
analyzing the process of formation of the dyadosphere during the process of gravitational collapse itself.
\end{itemize}

%INDEX%%%%%%%%%%%%%%%%%%%%%%%%%%%%%%%%%%%%%%%%%%%%%%%%%%%%%%%%%%%%%%%
\clearpage
\addcontentsline{toc}{section}{Index}
\flushbottom
\printindex
%%%%%%%%%%%%%%%%%%%%%%%%%%%%%%%%%%%%%%%%%%%%%%%%%%%%%%%%%%%%%%%%%%%%%

\end{document}